\pdfoutput=1
\documentclass[twocolumn,showpacs,preprintnumbers,amsmath,amssymb]{revtex4}

\usepackage{graphicx}
\usepackage{dcolumn}
\usepackage{bm}

\begin{document}


\title{Momentum distribution and non-Fermi liquid behaviour in low-doped two-orbital model: finite-size cluster quantum Monte Carlo approach}

\author{Vladimir A. Kashurnikov}
\affiliation{%
National Research Nuclear University MEPhI Kashirskoye shosse 31, Moscow, 115409, Russian Federation 
}%
\author{Andrey V. Krasavin}
\affiliation{%
National Research Nuclear University MEPhI Kashirskoye shosse 31, Moscow, 115409, Russian Federation 
}%
\author{Yaroslav V. Zhumagulov}
\affiliation{%
National Research Nuclear University MEPhI Kashirskoye shosse 31, Moscow, 115409, Russian Federation 
}%
\date{\today}

\begin{abstract}
The two-dimensional two-orbital Hubbard model is studied with the use of finite-size cluster  world-line quantum Monte Carlo algorithm. This model is widely used for simulation of the band structure of FeAs clusters, which are structure elements of Fe-based high-temperature superconductors.  The choice of a special basis set of hyper-sites allowed to take into account four-fermion operator terms and to overcome partly the sign problem. Spectral functions and the density of states for various parameters of the model are obtained in the undoped and low-doped regimes. The correlated distortion of the spectral density with the change of doping is observed, and the applicability of the "hard-band" approximation in the doped regime is demonstrated. Profiles of the momentum distribution are obtained for the first Brillouin zone; they have pronounced jump near the Fermi level, which decreases with the growth of the strength of the interaction. The invariance of the Fermi surface with respect to the strength of the interaction is testified. Nesting is found in the case of electron and hole doping. Fermi-liquid parameters of the model are derived. $Z$-factor grows sharply with the increasing the level of  doping, and monotonously decreases with the growth of the strength of the interaction. Moreover, electron-hole doping asymmetry of the $Z$-factor is revealed. The non-Fermi liquid behaviour and the deviation from Luttinger theorem are observed.
\end{abstract}

\maketitle

\section{Introduction}
\label{Introduction}
The role of electron correlations in high-temperature superconductors (HTSC) based on iron \cite{1} is decisive in the formation of physical properties of such systems. The effect of strong Coulomb interactions on the nature of superconductivity and the formation of a complex phase diagram including antiferromagnetic, structural and superconducting ordering, is a major focus of interest ~\cite{2,3}. Like copper HTSC, iron-based superconductors are characterized by a strongly expressed anisotropy, and have a structure of closely spaced atomic planes of Fe and As (for pnictides). The main contribution to the band structure near the Fermi surface is provided by two orbitals, $d_{xz}$  and $d_{yz}$. Therefore, the simplest model Hamiltonian for iron-based HTSC reflecting their crystal and electronic structure, is the two-dimensional generalized two-orbital Hubbard model \cite{4}, which has been intensively studied since the discovery of iron-based HTSC with the use of various approximate analytical and numerical methods. The mean-field approximation was used in \cite{5} to determine the magnetic order parameter and the electron density of states; the symmetry of the superconducting order parameter was investigated in \cite{6} using the random-phase approximation. Numerical methods generally also use various simplifications for modeling the Hamiltonian of the two-orbital model. Charge stripes states were obtained in \cite{7} within the real-space Hartree-Fock approximation away from the half-filling. Quasi one-dimensional "ladder" geometry was used in \cite{8} for the study of the properties of the ground state of the two-orbital model by density matrix renormalization group approach. In the recent paper \cite{9}, the scheme "Monte-Carlo + mean-field", which has successfully approved itself in study of one-dimensional models, was generalized, and thermal properties of the two-orbital model were investigated. Exact numerical simulations of the two-orbital model, which include the method of exact diagonalization and the quantum Monte Carlo algorithms, are also limited in their applications. The problem of the two-orbital model for pnictides was solved in full by the exact diagonalization technique in \cite{10}; however, the maximum size of the lattice, which one could managed to diagonalize, was $\sqrt{8} \times\sqrt{8}$.

The applicability of finite-size cluster quantum Monte Carlo algorithms that allow, in principle, to obtain exact results, is limited in this case by the sign problem, which exponentially slows calculations at a sufficiently low temperature and/or away from
the half-filling.

In this work, with choosing specific basis states for the Hamiltonian of the two-orbital model and the use of CTWL-algorithm \cite{11}, we were able to overcome partly the sign problem and to obtain data for Matsubara Green's function at sufficiently low temperature in a wide range of model parameters. Our previous studies for FeAs clusters of size from $3 \times 3$ to $10 \times 10$ have shown the possibility of effective $A_{1g}$-pairing \cite{11,13,14}. Note, that all calculations were performed within the limits of the \textit{full} two-orbital model. Accurate accounting of non-diagonal four-fermion terms $a_{i,\sigma}^{+}a_{j,\sigma^{'}}a_{l,\sigma^{''}}^{+}a_{k,\sigma^{'''}}$ which can play a key role in the formation of correlation properties, and which, from our point of view, can not be neglected, is the main feature of our algorithm.

 Using the relationship between Matsubara Green's function and the density of electron states, we have derived the momentum distribution and the Fermi liquid parameters of the model: quasi-particle weight ($Z$-factor), self-energy $\Sigma$, and scattering rate $\Gamma$, and studied the properties of the model under doping. The momentum distribution have a clear jump near the Fermi level; the value of this jump is sensitive to the strength of the interaction. The asymmetry in the behavior of $Z$-factor was observed at electron and hole doping, which reflects the asymmetry of the Hamiltonian. The evolution of the Fermi surface was studied at changing the model parameters and the level of doping; it is shown that the “hard-band approximation” is acceptable for the analysis of the topology of the Fermi surface at low doping and finite value of the interaction. An increase of the volume of the Fermi surface with the growth of interaction was observed, which shows the deviation from Luttinger theorem.

The paper is organized as follows. In the next section we introduce the Hamiltonian of the two-dimensional two-orbital model for pnictides. In Section III, the relation between Matsubara Green’s function and the momentum distribution is demonstrated. Section IV is devoted to the study of the half-filling case of the model. In Section V, the spectral function and the total density of states are restored, and their dependences on the parameters of the model are obtained. The doped regime is studied in Section VI, and the Fermi-liquid parameters of the model are derived in Section VII. In Section VIII we summarize the results obtained.

\section{The model}
\label{2}
First introduced in \cite{4}, the two-orbital model for iron-based HTSC takes into account the real crystal structure of these compounds, as well as the two-dimensional nature of physical properties. Provided that the main contribution to the formation of the band structure near the Fermi level make $3d$-states of iron atoms \cite{10,12}, the Hamiltonian of the model is presented as follows:

\begin{widetext}
\begin{equation}\label{eq:Hamiltonian}
H=H_{int}+H_{kin}
\end{equation}
\begin{eqnarray}
H_{int}=U\sum_{\boldsymbol{i}\alpha }{n_{\boldsymbol{i}\alpha \uparrow }n_{\boldsymbol{i}\alpha \downarrow }}+V\sum_{\boldsymbol{i}}{n_{\boldsymbol{i}x}n_{\boldsymbol{i}y}}-\mu \sum_{\boldsymbol{i}}{n_{\boldsymbol{i}}}-J\sum_{\boldsymbol{i}}{\left(n_{\boldsymbol{i}x\uparrow }n_{\boldsymbol{i}y\uparrow }+n_{\boldsymbol{i}x\downarrow }n_{\boldsymbol{i}y\downarrow }\right)}-\nonumber \\
-J\sum_{\boldsymbol{i}}{\left(a^+_{\boldsymbol{i}x\downarrow }a_{\boldsymbol{i}x\uparrow }a^+_{\boldsymbol{i}y\uparrow }a_{\boldsymbol{i}y\downarrow }+a^+_{\boldsymbol{i}x\uparrow }a_{\boldsymbol{i}x\downarrow }a^+_{\boldsymbol{i}y\downarrow }a_{\boldsymbol{i}y\uparrow }+a^+_{\boldsymbol{i}x\uparrow }a_{\boldsymbol{i}y\downarrow }a^+_{\boldsymbol{i}x\downarrow }a_{\boldsymbol{i}y\uparrow }+a^+_{\boldsymbol{i}y\uparrow }a_{\boldsymbol{i}x\downarrow }a^+_{\boldsymbol{i}y\downarrow }a_{\boldsymbol{i}x\uparrow }\right)}
\nonumber
\end{eqnarray}
\begin{eqnarray}
H_{kin}=-t_1\sum_{\boldsymbol{i}\sigma }{\left(a^+_{\boldsymbol{i}x\sigma }a_{\boldsymbol{i}\boldsymbol{+}\boldsymbol{x},x\sigma }+a^+_{\boldsymbol{i}y\sigma }a_{\boldsymbol{i}\boldsymbol{+}\boldsymbol{y},y\sigma }\right)}-t_2\sum_{\boldsymbol{i}\sigma }{\left(a^+_{\boldsymbol{i}y\sigma }a_{\boldsymbol{i}\boldsymbol{+}\boldsymbol{x},y\sigma }+a^+_{\boldsymbol{i}x\sigma }a_{\boldsymbol{i}\boldsymbol{+}\boldsymbol{y},x\sigma }\right)}-\nonumber \\
-t_3\sum_{\boldsymbol{i}\sigma }{\left(a^+_{\boldsymbol{i}x\sigma }a_{\boldsymbol{i}\boldsymbol{+}\boldsymbol{x}\boldsymbol{+}\boldsymbol{y},x\sigma }+a^+_{\boldsymbol{i}x\sigma }a_{\boldsymbol{i}\boldsymbol{+}\boldsymbol{x}\boldsymbol{-}\boldsymbol{y},x\sigma }+a^+_{\boldsymbol{i}y\sigma }a_{\boldsymbol{i}\boldsymbol{+}\boldsymbol{x}\boldsymbol{+}\boldsymbol{y},y\sigma }+a^+_{\boldsymbol{i}y\sigma }a_{\boldsymbol{i}\boldsymbol{+}\boldsymbol{x}\boldsymbol{-}\boldsymbol{y},y\sigma }\right)}+\nonumber \\
+t_4\sum_{\boldsymbol{i}\sigma }{\left(a^+_{\boldsymbol{i}x\sigma }a_{\boldsymbol{i}\boldsymbol{+}\boldsymbol{x}\boldsymbol{-}\boldsymbol{y},y\sigma }+a^+_{\boldsymbol{i}y\sigma }a_{\boldsymbol{i}\boldsymbol{+}\boldsymbol{x}\boldsymbol{-}\boldsymbol{y},x\sigma }-a^+_{\boldsymbol{i}x\sigma }a_{\boldsymbol{i}\boldsymbol{+}\boldsymbol{x}\boldsymbol{+}\boldsymbol{y},y\sigma }-a^+_{\boldsymbol{i}y\sigma }a_{\boldsymbol{i}\boldsymbol{+}\boldsymbol{x}\boldsymbol{+}\boldsymbol{y},x\sigma }\right)}+h.c.\nonumber
\end{eqnarray}
\end{widetext}
Here operator $a^+_{\boldsymbol{i}x\left(y\right)\sigma }(a_{\boldsymbol{i}x(y)\sigma })$ creates (annihilates) an electron with spin projection $\sigma $ on site $\boldsymbol{i}$\textbf{ }and orbital $x(y)$; $t_i,\ i=1,\dots ,4$ are hopping amplitudes between orbitals $x(y)$; $U$, $V$, and $J$ are Coulomb interaction terms; $\mu $ is the chemical potential.
In this work we used the quantum continuous time world-line Monte Carlo algorithm (CTWL-algorithm), adapted for the two-orbital model \cite{11}. The algorithm is free from Wick’s decomposition and allows the calculation of the Matsubara Green's function and obtaining information about the quasiparticle spectrum and its dependence on temperature and interaction parameters.

According to our studies \cite{11,13}, the exchange term is important for the correct description of the correlation properties of the system and the manifestation of a certain type of pairing symmetry. At the same time, this term is what distinguishes the model \eqref{eq:Hamiltonian} from the usual generalized Hubbard model, and leads to difficulties in the implementation of the simulation by quantum Monte Carlo algorithm. Each term of the Hamiltonian \eqref{eq:Hamiltonian} should be considered separately for inclusion in the scheme of the algorithm. 
All the features of the numerical modeling, encoding of the basic states, and calculation of correlation functions and Green's function in the framework of the CTWL algorithm are described in detail in \cite{11,13,14}.

The parameters of $H_{kin}$  in \eqref{eq:Hamiltonian} were taken the same as in \cite{10}:
\begin{equation}
t_1=0.058(eV); t_2=0.22; t_3=-0.21 ;t_4=-0.08;
\end{equation}
and the relationships between Coulomb terms in  $H_{int}$  are the following:
\begin{equation}\label{eq:interaction}
V=0.5U; J=0.25U.
\end{equation}

\section{Matsubara Green's function and momentum distribution}
\label{3}
The Matsubara Green's function
\begin{equation}
G_{ij\sigma\sigma^{'}}(\tau )=- \left\langle T_{\tau }a_{\boldsymbol{i}\alpha \sigma }(\tau )a^+_{\boldsymbol{j}\beta {\sigma }^{'}}(0) \right\rangle
\end{equation}
was calculated for clusters up to $10\times 10$ FeAs-cells in the temperature range $1/T=1\div 5$ for various values of $U=1\div 8$ while maintaining the relation (\ref{eq:interaction}). An off-diagonal term of the form
\begin{equation}
-\gamma \sum_{ij\alpha }{\left(a^+_{i\alpha \sigma }+a_{i\alpha \sigma }\right)}
\end{equation}
has been added in the calculation. Here $\gamma \sim 5\times 10^{-3}$. On the one hand, such addition to the Hamiltonian does not modify the calculation results within the accuracy achieved, and on the other hand, the input of such controlled off-diagonal term substantially "animates" the statistics and increases the speed of convergence. In addition, the presence of this term allows directly accumulate statistics on Matsubara Green's function in the world-line algorithm.

Actually, the momentum distribution $n(\boldsymbol{k})$ can be obtained from the relation 
\begin{eqnarray}\label{eq:direct_approximation}
n\left(\boldsymbol{k}\right)=G\left(\boldsymbol{k},\tau \to -0\right);\nonumber\\ 
G\left(\boldsymbol{k},\tau \right)=\sum_{ij}{G_{ij}\left(\tau \right)}e^{i\boldsymbol{k}{\boldsymbol{r}}_{ij}}.
\end{eqnarray}
Direct approximation \eqref{eq:direct_approximation} in a numerical calculation leads to uncontrolled errors; therefore, we have used the following expression:
\begin{equation}\label{eq:n}
n_\sigma\left(\boldsymbol{k}\right)=\int{\frac{A_{\sigma }\left(\boldsymbol{k},\omega \right)}{1+e^{\beta (\omega -\mu )}}}d\omega.
\end{equation}

Normalization of $n_\sigma(\boldsymbol{k})$  was made on the average occupation numbers of orbitals calculated independently in the same calculation for a given chemical potential. To calculate \eqref{eq:n}, the spectral function $A_{\sigma }(\boldsymbol{k},\omega )$ and the total density of states $N\left(\omega \right)=\sum_{\boldsymbol{k}\sigma }{A_{\sigma }(\boldsymbol{k},\omega )}$ were recovered from the integral equation relating the spectral function with Matsubara Green's function,
\begin{equation}
G_{\sigma }\left(\boldsymbol{k},\tau \right)=-\int{\frac{A_{\sigma }\left(\boldsymbol{k},\omega \right)e^{-\tau \left(\omega -\mu \right)}}{1+e^{-\beta \left(\omega -\mu \right)}}}d\omega,
\end{equation}
with the use of a stochastic procedure, the details of which are explained in Section V.

\section{The half-filling case}
\label{4}
It should be noted that the results for clusters with the size of $6\times 6$  FeAs-cells and larger were no longer depended on the system size indicating the applicability of the results for the analysis of the properties of a macroscopic system. Further data are presented primarily for $8\times 8$  clusters.

\begin{figure}
\includegraphics[width=8cm]{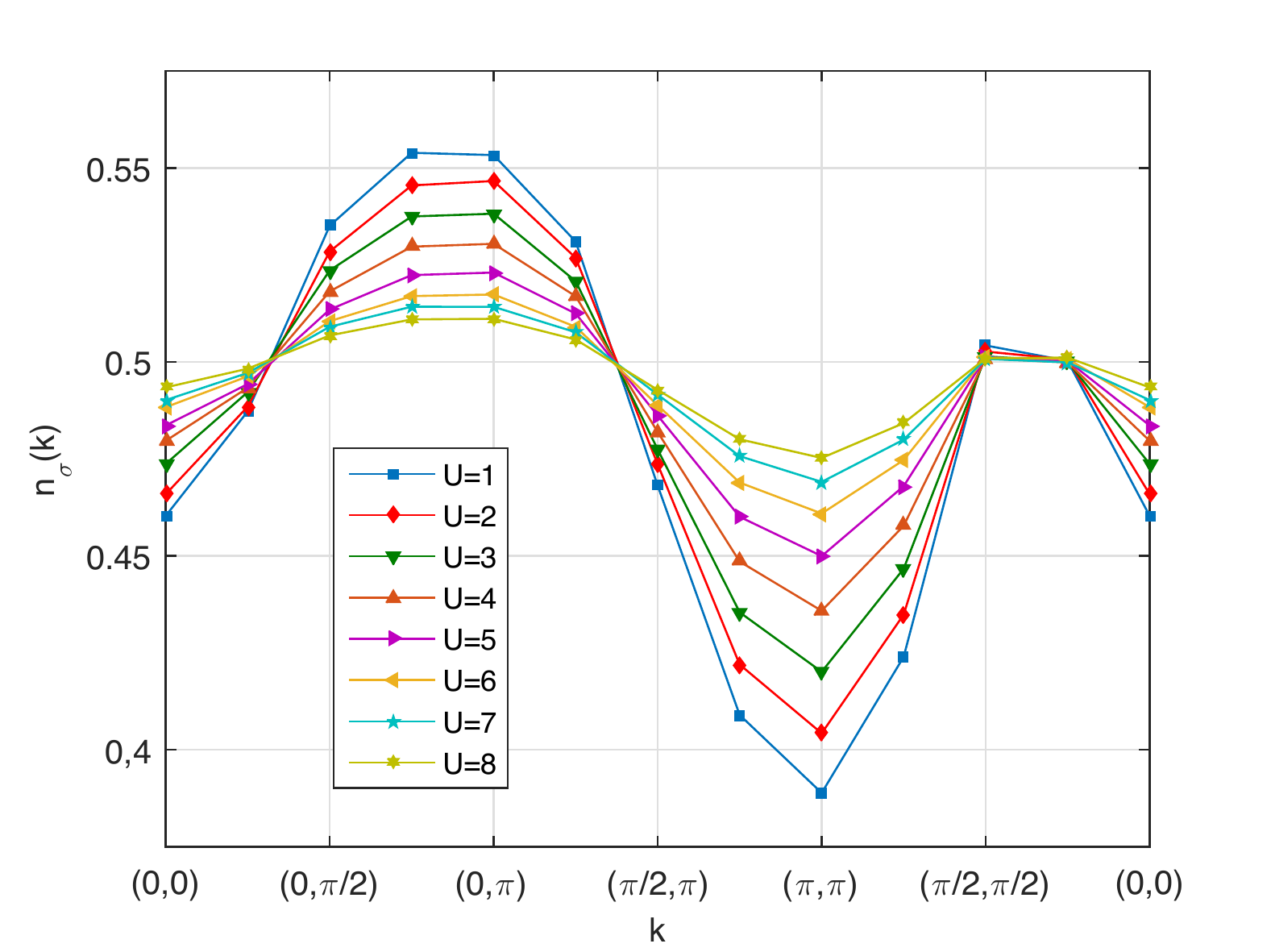}
\caption{\label{fig:MD} (Color online) Profiles of the momentum distribution along the main crystallographic directions, depending on the parameter $U$ at half-filling. Cluster $8 \times 8$; $\beta$ = 1}
\end{figure}

Figure \ref{fig:MD} shows the profiles of the momentum distribution depending on the parameter $U$ at half-filling along the main crystallographic directions. The data are presented for the inverse temperature $\beta =1$. This is a reasonably low temperature as $\beta t_{max}=\beta t_2\sim 0.2$; in addition, the sign problem did not allow lowering the temperature significantly. We were not able to investigate the features of superconducting correlations at that temperature, but succeeded in obtaining the overall picture of the band structure and the momentum distribution.

The Fermi level was determined by half-filling of the distribution (see \cite{15} for the peculiarities of determining the Fermi surface (FS) at nonzero interaction and $T>0$). In the graphs, it is seen as a rather abrupt change of occupation numbers, and, in a first approximation, it coincides with the maximum gradient of the momentum distribution (Fig. \ref{fig:MD}). There are no distinct filled and unfilled bands; the interaction significantly blurs the profile. The more strong is the interaction, the more blurred is the jump, and the more flattened is the distribution.
It is worthy to note that in a first approximation, the FS is independent of the strength of the interaction; all the curves intersect at the same points of the Brillouin zone. This invariance of the FS for FeAs-systems was reported in \cite{16}, and, in addition, was demonstrated in the framework of the extended Hubbard model \cite{17}. However, a more detailed examination of the data in Fig. \ref{fig:MD} shows that the distribution curves for various interaction parameters cross the half-filling level at slightly different momenta, indicating a weak dependence of the shape of the FS on interaction.

\begin{figure}
\includegraphics[width=8cm]{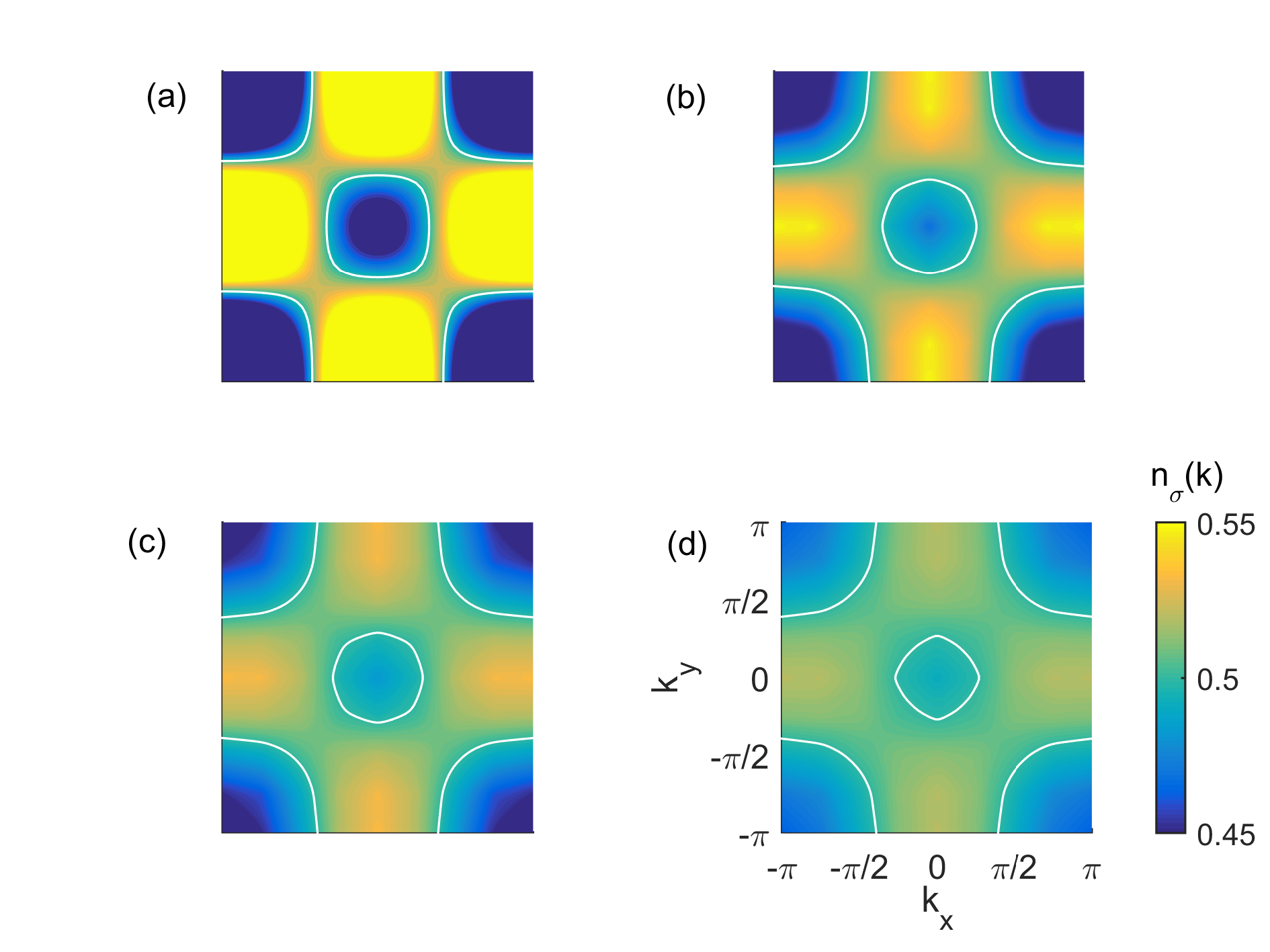}
\caption{\label{fig:FS-1} (Color online) Momentum distribution in the first Brillouin zone and the Fermi surface (white line) for various values of $U$. Cluster $8 \times 8$; $\beta = 1$; half-filling. a) $U= 0$; b) $ U = 2$; c) $U = 4$; d) $ U = 8$} 
\end{figure}

The momentum distribution for the first Brillouin zone is presented in Fig. \ref{fig:FS-1} for various interaction parameters. The FS is shown in white and corresponds to the level of half-filling $n(\boldsymbol{k}) = 1.0$. Its shape is generally consistent with the results of ARPES experiments and numerical calculations \cite{16,18,19,20,21,22}. A typical picture of hole pockets at the points $\boldsymbol{\Gamma}$ and \textbf{M} is visible, which was observed also in LnOFeAs (1111) and $\text{BaFe}_2\text{As}_2$ (122) \cite{16}. A similar picture was observed in LiFeAs \cite{21, 22}.

As can be seen from Fig. \ref{fig:FS-1}, the form of the FS is weakly dependent on the interaction; with increasing of $U$ a slight redistribution of excitations occurs from the hole to the electronic branch, simultaneously with the reducing of the hole pockets and the flattening of the momentum distribution (Fig. \ref{fig:MD}). The shrinkage of the electron and hole subbands is accompanied by a tendency of the straightening of the FS. This is particularly evident around the hole pocket in the center of the zone, at the point $\boldsymbol{\Gamma}$: with the increase of the interaction the boundary of the FS around the shrinking hole pocket turns into a diamond; and a tendency to nesting is observed.

\begin{figure}
\includegraphics[width=8cm]{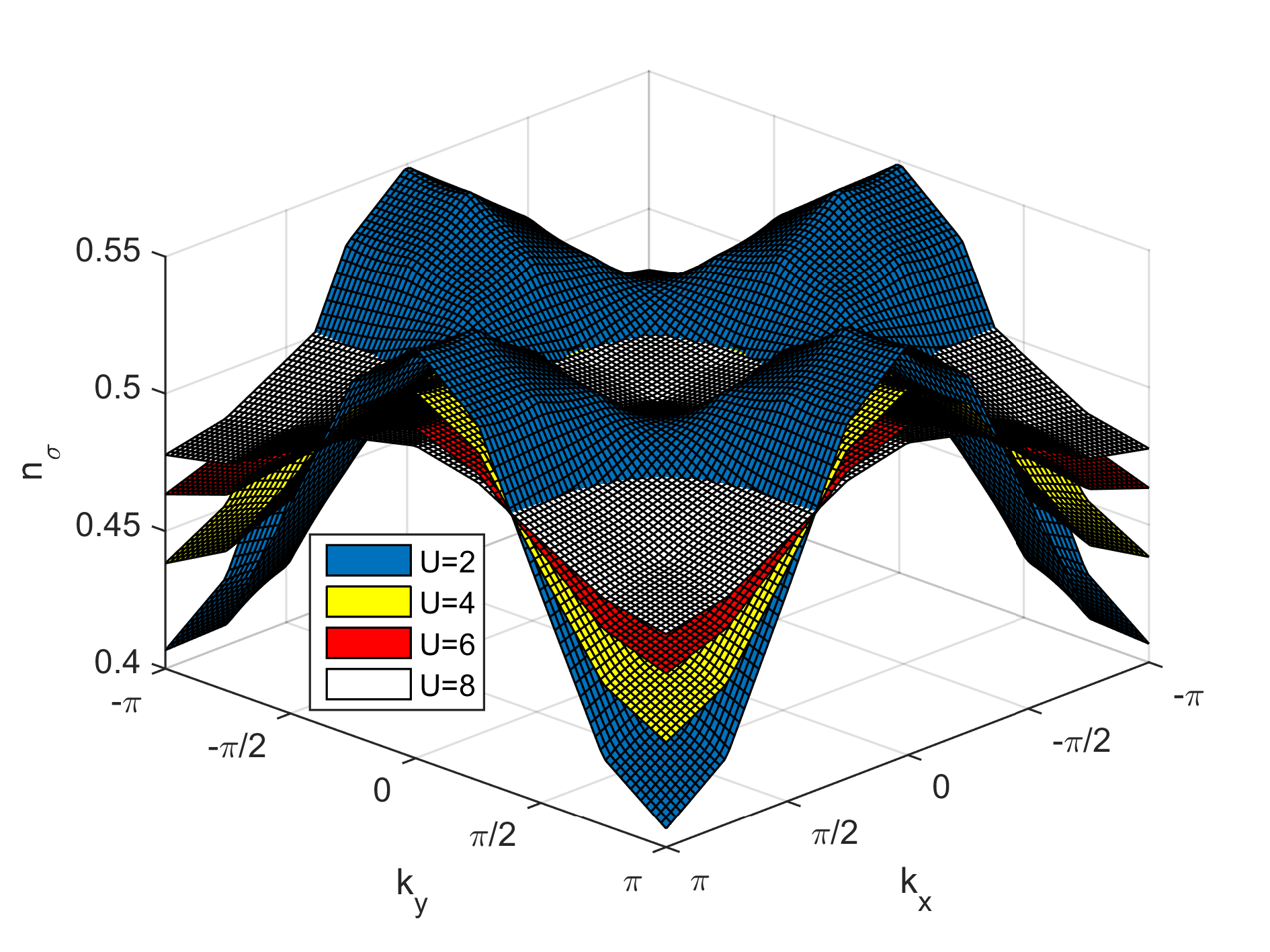}
\caption{\label{fig:FS-2} (Color online) Momentum distribution in the first Brillouin zone as a function on the interaction parameter $U$. Cluster $8 \times 8$; $\beta = 1$; half-filling} 
\end{figure}

The features of the momentum distribution are also clearly visible in 3D-picture shown in Fig. \ref{fig:FS-2}: the flattening of the distribution with increase of the interaction, as well as the presence of the hole pockets at the center ($\boldsymbol{\Gamma}$) and the periphery (\textbf{M}) of the zone, the filling of which increases with $U$.

\section{The density of states}
\label{5}
The spectral function $A_{\sigma }\left(\boldsymbol{k},\omega \right)$ is related to the Matsubara Green's function by the integral equation
\begin{equation}\label{eq:integral}
G_{\sigma }\left(\boldsymbol{k},\tau \right)=-\int{\frac{A_{\sigma }\left(\boldsymbol{k},\omega \right)e^{-\tau \omega }}{1+e^{-\beta \omega }}d\omega},
\end{equation}
where $G\left(\boldsymbol{k},\tau \right)\equiv -\left\langle T_{\tau }\left[a_{\boldsymbol{k}}(\tau )a^{+}_{\boldsymbol{k}}(0)\right]\right\rangle $; $\boldsymbol{k}$ runs the full set of allowed momenta in the system; $\tau $ is imaginary time; $\sigma $ is a spin projection. The task of restoring the spectral function from the equation \eqref{eq:integral}, therefore, is the ill-posed problem of solving the Fredholm integral equation of the first kind. 

One of the most effective schemes for solving the equation \eqref{eq:integral} was proposed in \cite{23}. The main idea of the method is to approximate the density of states $A\left(\omega \right)$ by a piecewise constant function $\widetilde{\rho }\left(\omega \right)$,
\begin{equation}
A\left(\omega \right)\longleftarrow \widetilde{\rho }\left(\omega \right)=\sum{{\chi }_{c,w,h}\left(\omega \right)}
\end{equation}
representing the sum of rectangles determined by the center position $c$, the width $w$ and the height $h$. The approximated Matsubara Green's function ${\tilde{G}}_{\chi }\left(\tau \right)$ in this case is
\begin{eqnarray}\label{eq:green_approximation}
{\tilde{G}}_{\chi }\left(\tau \right)=\int^{+\infty }_{-\infty }{\frac{e^{-\tau \omega }}{1+e^{-\beta \omega }}}\widetilde{\rho }\left(\omega \right)d\omega =\nonumber \\
=\left.\frac{e^{-(\beta -\tau )\omega }}{\beta }{{}^{\ }_2F}_1\left(1,1-\frac{\tau }{\beta },2-\frac{\tau }{\beta },-e^{\beta \omega }\right)\right|\genfrac{}{}{0pt}{}{c+\frac{w}{2}}{c-\frac{w}{2}},
\end{eqnarray}
where $_2F_1$ is the hypergeometric function.

The algorithm is based on minimizing the deviation between the input Matsubara Green's function (obtained from finite-size cluster quantum Monte Carlo calculations) and the approximated function \eqref{eq:green_approximation} by generating stochastic configurations consisting of various numbers of rectangles ${\chi }_{c,w,h}$. Various functionals can be used for the calculation of the deviation; in this work good convergence was achieved with the use of the least squares functional
\begin{equation}
\Delta_G=\int^{\beta }_{0}{\left(G\left(\tau \right)-\tilde{G}\left(\tau \right)\right)}^2d\tau.
\end{equation}

\begin{figure}
\includegraphics[width=8cm]{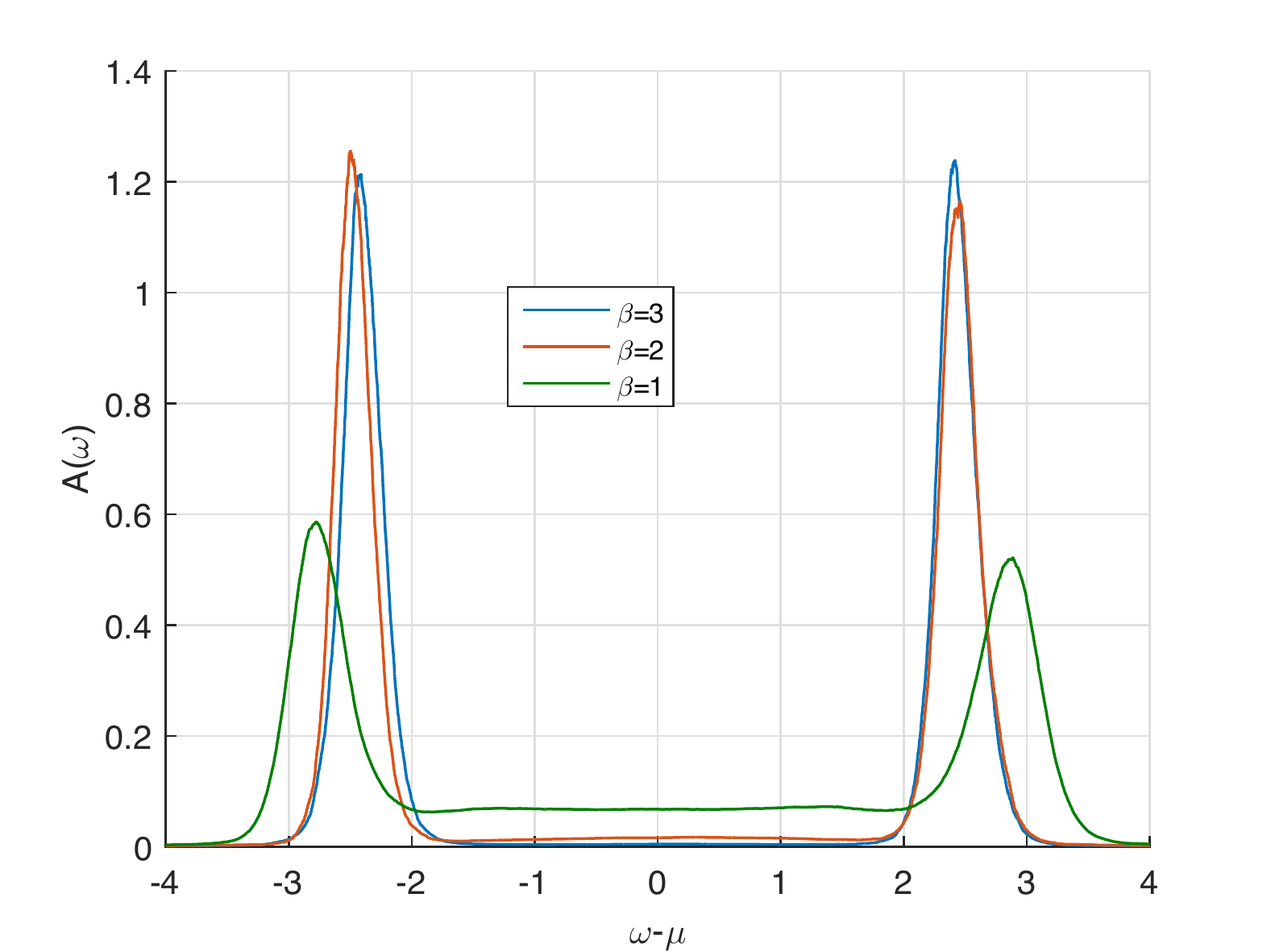}
\caption{\label{fig:DOS-1} (Color online) The density of states at various temperatures. $U=4$; half-filling} 
\end{figure}	

Figure \ref{fig:DOS-1} shows the total density of states (DOS) depending on temperature at half-filling, i.e. in the undoped regime. The significant variation of the $\beta =1$ curve from the others is presumably due to the temperature effect; it can be concluded that the temperature $\beta >1$ is sufficiently low for the study.

\begin{figure}
\includegraphics[width=8cm]{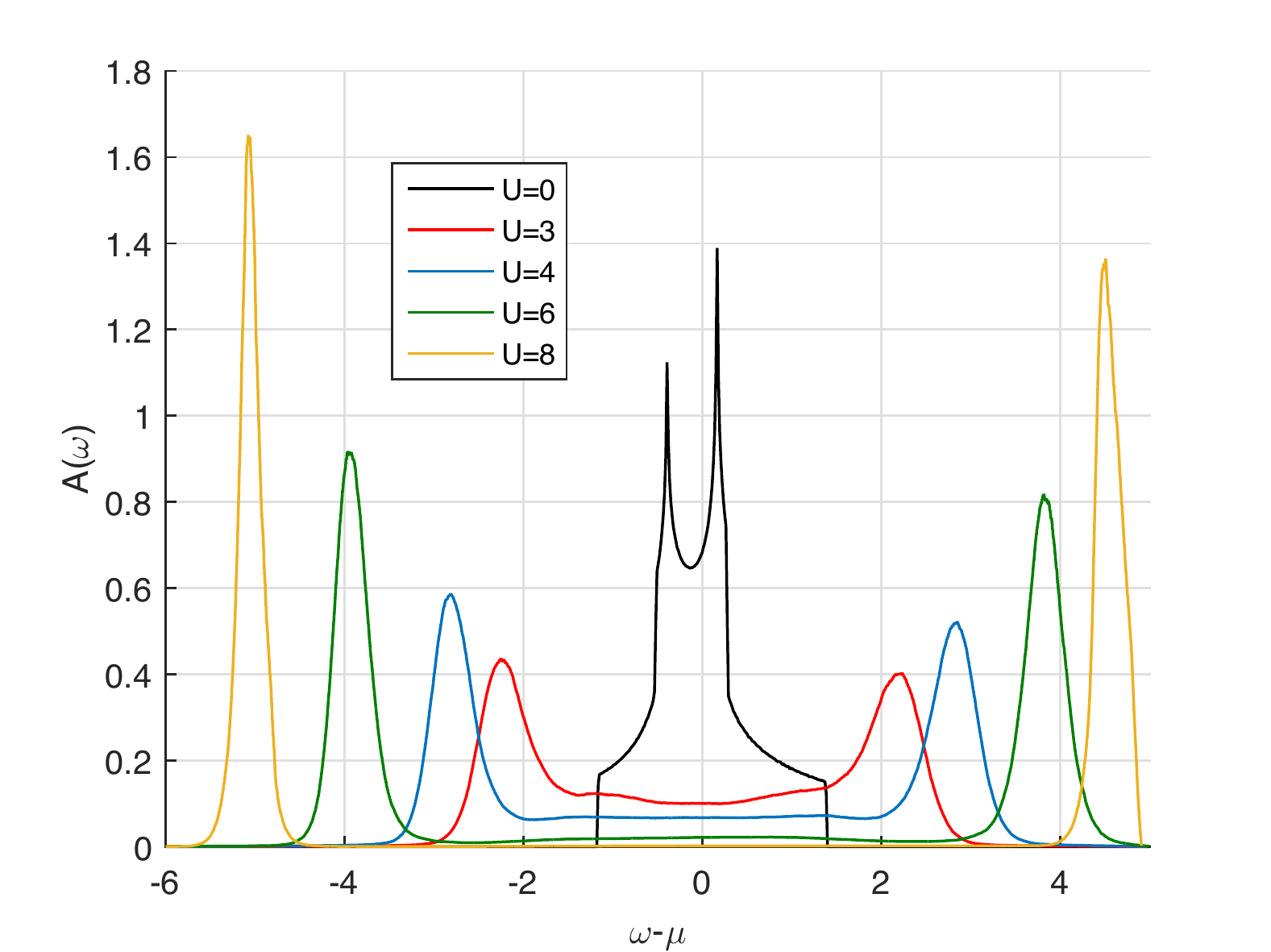}
\caption{\label{fig:DOS-2} (Color online) The total DOS as a function of $U$. Cluster $8\times 8$, $\beta=1$. The DOS for a free system is obtained analytically and is shown in black} 
\end{figure}	

The total DOS depending on the value of the interaction is shown in Fig. \ref{fig:DOS-2} at the half-filling. The DOS in the absence of interaction (black line) was calculated analytically for comparison. The calculations were performed at $\beta =1$, so a thermal broadening of the bands takes place; however, we believe that it should not prevent to see the evolution of the bands with changing of $U$. This choice of temperature is associated with the convergence of quantum Monte Carlo algorithm, namely, with the sign problem. Only at this temperature we were able to perform the calculations at a series of values of the interaction parameters with sufficient accuracy.

The interval between the bands increases with the increase of the interaction, and is close to the value of $U$, but not identical with it, since the interaction part of the Hamiltonian \eqref{eq:Hamiltonian} is more complicated than the conventional Hubbard term. With the growth of the interaction the bands shrink turning into narrow peaks; this leads to a reduction of the dispersion and flattening of the momentum distribution as was noted in Section IV.

\begin{figure}
\includegraphics[width=9cm]{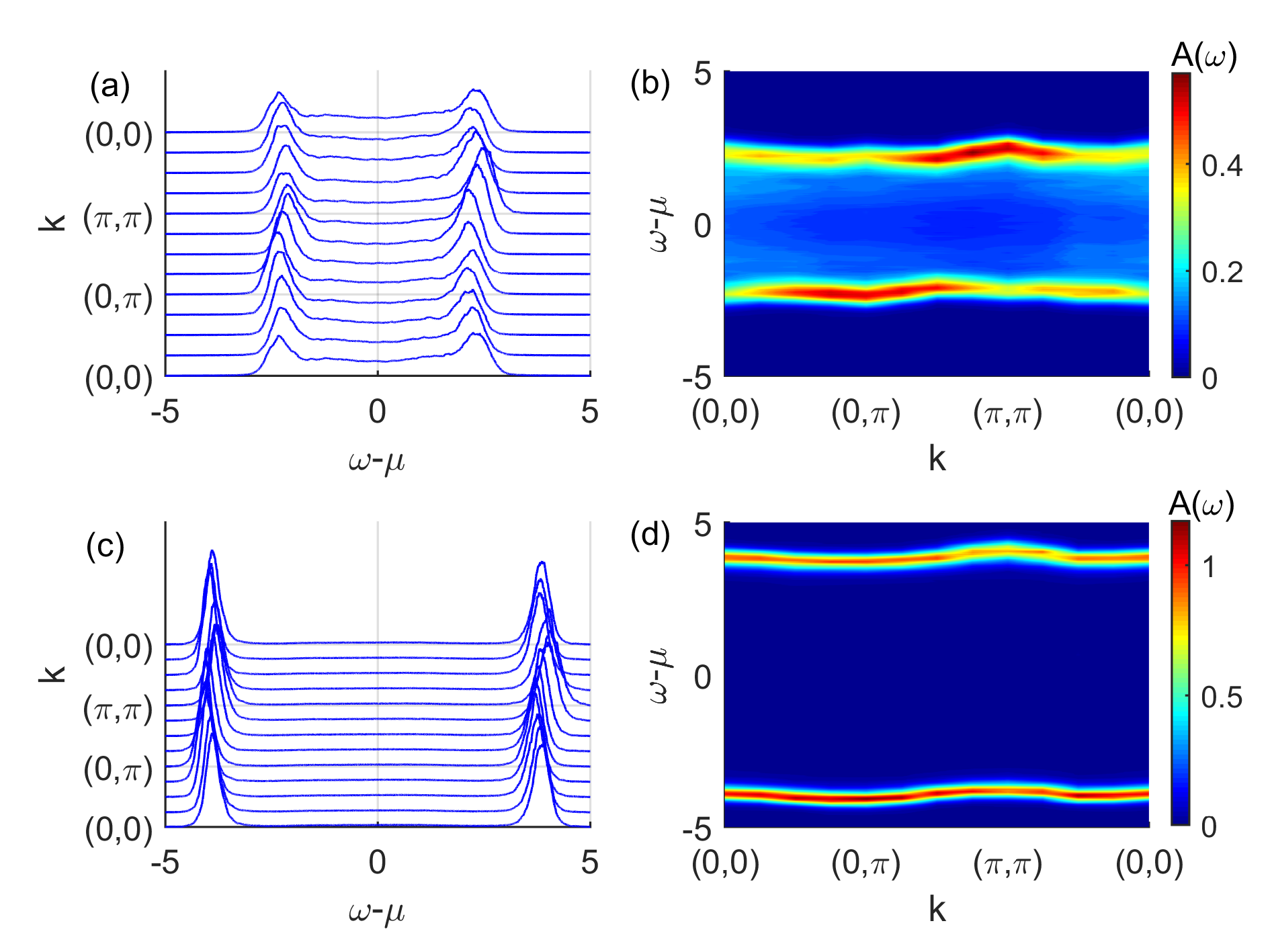}
\centering
\caption{\label{fig:DISP} (Color online) The spectral function along the main crystallographic directions. Left: profiles of the spectral function; right: the dispersion. a), b) $U=3$; c), d) $U=6$. Cluster $8\times 8$, $\beta=1$} 
\end{figure}

A clear picture of the flattening of the dispersion and shrinking of the bands is also visible in the spectral function and dispersion of excitations, and is shown in Fig. \ref{fig:DISP}.
 
Note that as the strength of the interaction increases, the quasiparticle approximation becomes more and more valid for electron and hole excitations, since the half-width of the spectral peaks (and hence the damping) decreases.

Consider now the doped case. As noted above, the finite-size cluster quantum Monte Carlo algorithm suffers from the minus-sign problem when calculating a two-dimensional Fermi system. This leads to the fact that the deviation from the half-filling may not afford calculation of thermodynamic average values, including the Green's function, with the reasonable accuracy. We have managed to obtain a good convergence in the doped case only at several parameters of $U$ and at temperatures not lower than $\beta =2$. The same problems were mentioned in \cite{24} where the doped single-band Hubbard model was studied.

\begin{figure}
\includegraphics[width=8cm]{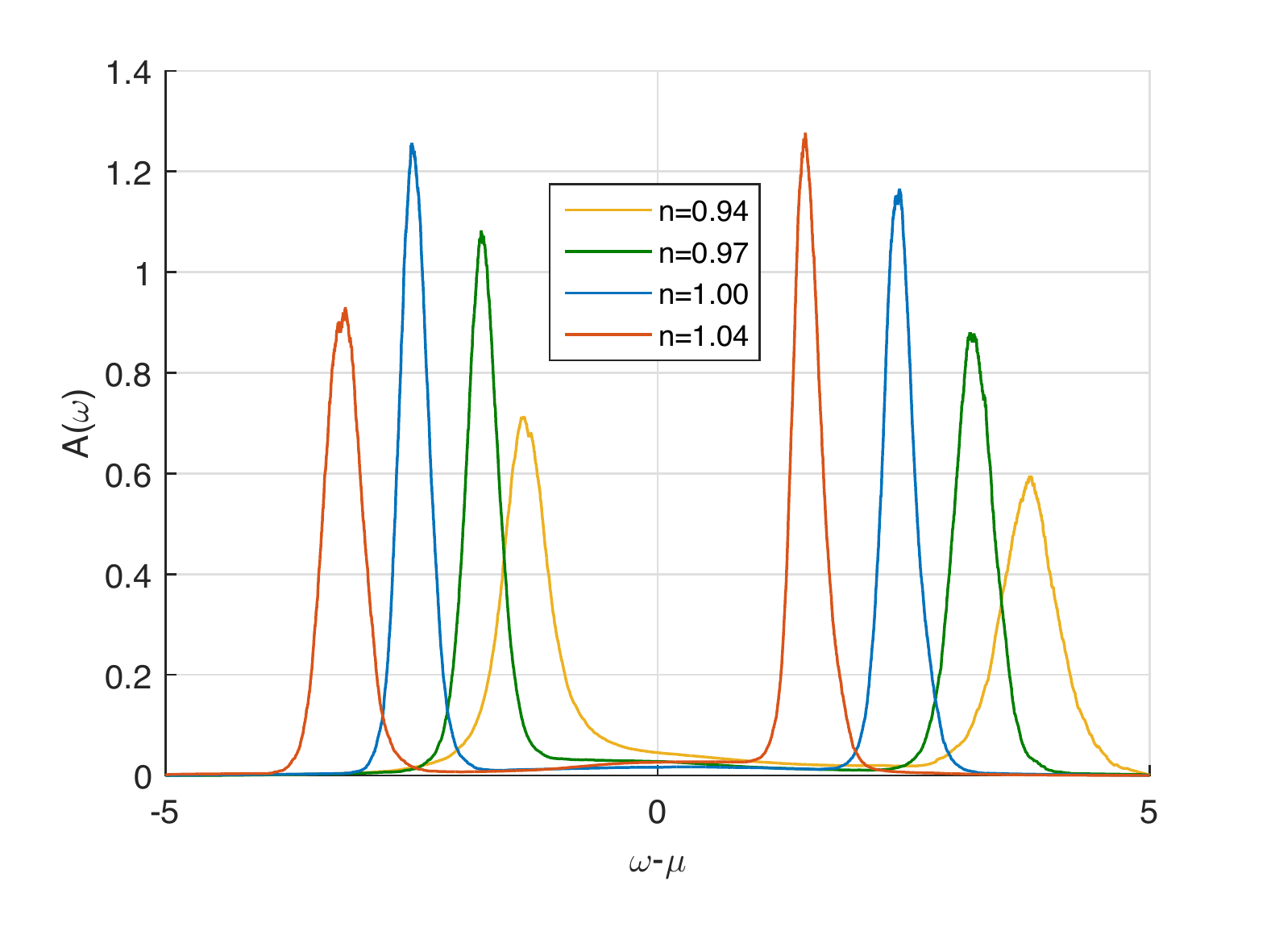}
\caption{\label{fig:DOS-doped} (Color online) The evolution of the density of states with change of doping near the half-filling. $U=4$;$\beta =2$} 
\end{figure}	

Figure \ref{fig:DOS-doped} shows the total density of states at a low level of electron and hole doping. The shape of the bands changes as the chemical potential moves from the electron to the hole doping region. Thus, at $n>1$ (electron doping) the right peak (unoccupied states) is higher, while at $n<1$ the left peak (filled states) becomes higher, as some redistribution of the spectral density occurs. The broadening of the bands is observed with increasing the level of doping (both hole and electron). The value of DOS at the Fermi level increases slightly with increasing the hole doping.

Therefore, the change of the number of carriers in the system is not equivalent to a simple filling of the bands (the so-called "hard-band approximation"), i.e., to a simple shift of the chemical potential at a constant density of states, which is the contribution of Coulomb correlations in this typical generalized Hubbard model. Similar correlation effects were also observed in \cite{24} for the Hubbard model in the low-doped regime. However, it will be shown further that the hard-band approximation can be used as a first approximation for the study of the Fermi surface topology and characteristics of the quasiparticle spectrum.

\section{The momentum distribution in the doped case}
\label{6}
\begin{figure}
\includegraphics[width=8cm]{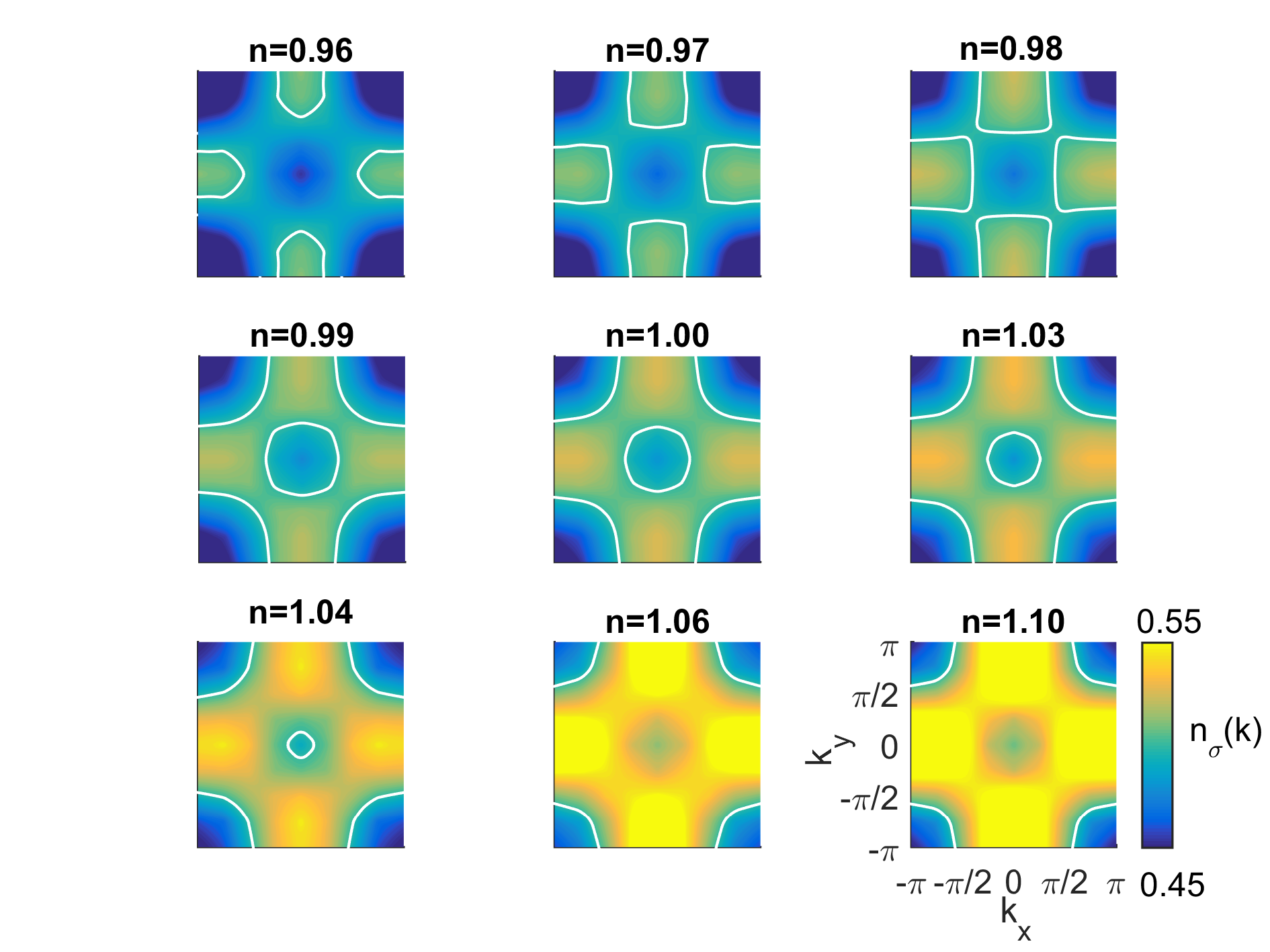}
\caption{\label{fig:FS-doped} (Color online) Momentum distribution and the FS (shown in white) as functions of the level of doping. Cluster $8\times 8$. $U=4$;$\beta =2$} 
\end{figure}

Figure \ref{fig:FS-doped} shows the profiles of the FS at $U=4$, $\beta =2$ in the doped case.

The first thing to note is the presence of nesting at electron doping and finite value of $U$, which is in agreement with ARPES experiments \cite{21, 22}. The changes of the Fermi surface are similar to those observed experimentally, for example, in $\text{LiFe}_{1-\text{x}}\text{Co}_{\text{x}}\text{As}$ \cite{22}. This is seen in the contour of the FS around the central hole area ($\boldsymbol{\Gamma}$), as in the case of the electron doping a diamond emerges (shrinking with the increase of the interaction) at an angle of 45 degrees to the edge of the Brillouin zone. In addition, the shape of the FS is different in this area for hole and electron doping. Indeed, the shape of a square oriented along the main crystallographic directions is almost retained under hole doping, i.e. nesting is observed (as shown by calculation, for any interaction).

It is useful to compare these data of the direct MC calculations with the hard-band approximation, or with a simple filling of the bands by the shift of the chemical potential with the use of the exactly calculated spectral density at the half-filling. This description of a strongly correlated system, as was shown in Section V, is not entirely correct, but considering the fact that the profile of the FS in a first approximation is weakly dependent on the interaction near the half-filling, it is possible to obtain data that will not be too much different from the accurate calculation. In addition, due to the better convergence of the Monte Carlo algorithm at the half-filling, a wider range of the values of $U$ can be studied.

\begin{figure}
\includegraphics[width=8cm]{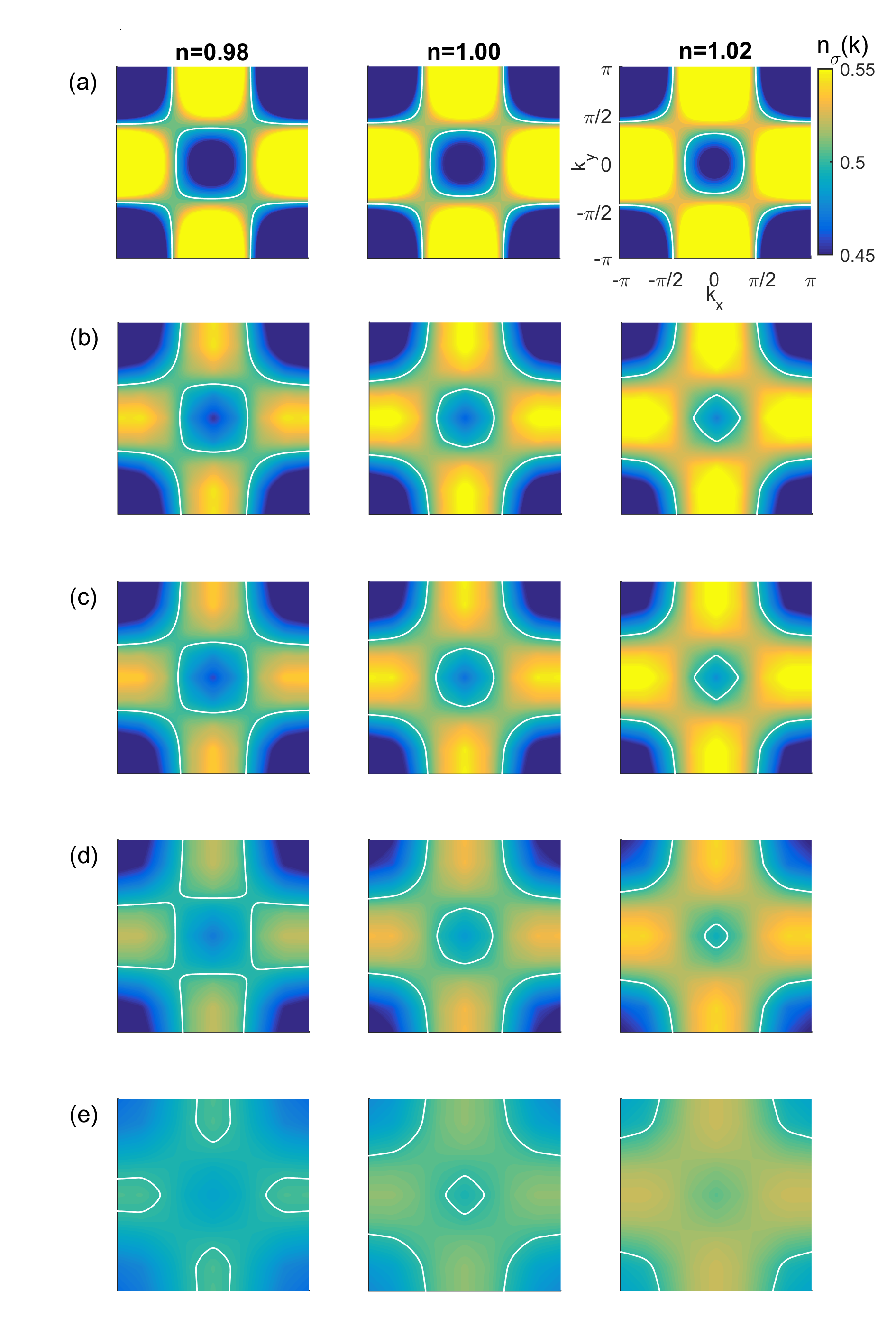}
\caption{\label{fig:FS-doped-approximation} (Color online) The distribution of occupation numbers in the hard-band approximation. a) $U=0$; b) $U=1$; c) $U=2$; d) $U=4$; e) $U=8$} 
\end{figure}

Figure \ref{fig:FS-doped-approximation} presents the momentum distribution in the hard-band approximation for various fillings and interaction parameters for $8\times 8$ cluster; for clarity, the analytical result for $U=0$ is also shown.

As well as in the exact calculation, nesting is observed here at electron doping and finite interaction; the square shape of the FS maintains at hole doping for any interaction. It can be seen that in the case of hole doping carriers tend to localize near the regions $(0,\pm \pi )$, $(\pm \pi ,0)$. Under electron doping the central hole pocket dramatically shrinks, and the curvature of the FS at the corners of the zone (\textbf{M}) is close to a circle, somewhat flattened only at sufficiently large interaction. Therefore, in a first approximation, the invariance of this section of the FS is preserved even for deviation from the half-filling.

Thus, the main features of the momentum distribution in the hard-band approximation are not much different from these of the exact calculation.

\begin{figure}
\includegraphics[width=8cm]{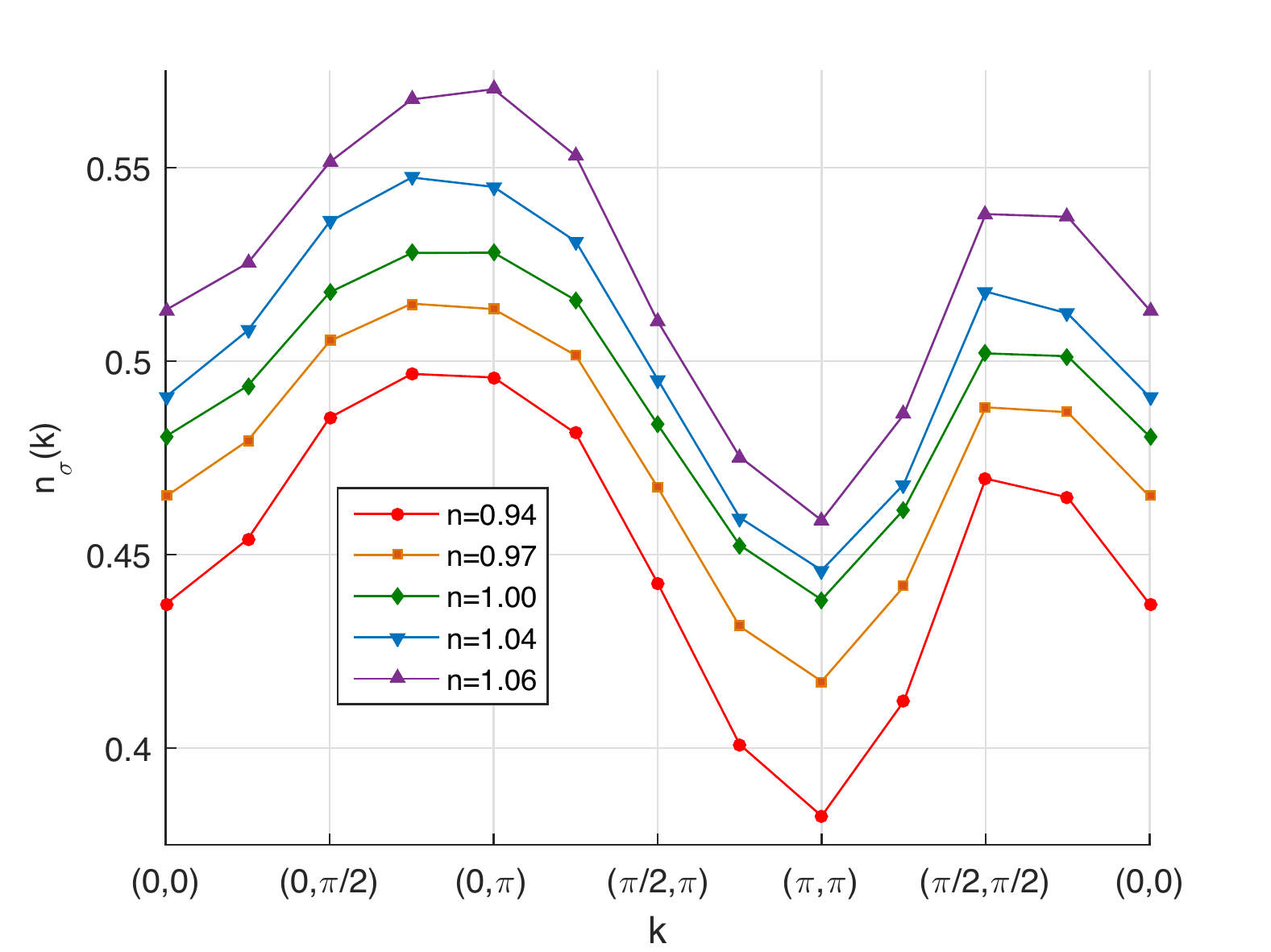}
\caption{\label{fig:MD-doped} (Color online) Momentum distribution for various levels of doping. $U=4$;$\beta =2$} 
\end{figure}	

Figure \ref{fig:MD-doped} shows the profiles of the momentum distribution along the main crystallographic directions in the first Brillouin zone for the same values of doping, as in Fig. \ref{fig:DOS-doped}. The following can be noted. The profiles with different fillings are substantially equidistant, which means that the bands are filled with virtually no distortion as the concentration changes, and the hard-band approximation works well. The momentum distributions vary little with the change of the filling, which is consistent with the statement that the momentum distribution of quasiparticles has the same features (including the Fermi jump) as the total momentum distribution. Interaction blurs distribution gradients as well as in the half-filling case.

\begin{figure}
\includegraphics[width=8cm]{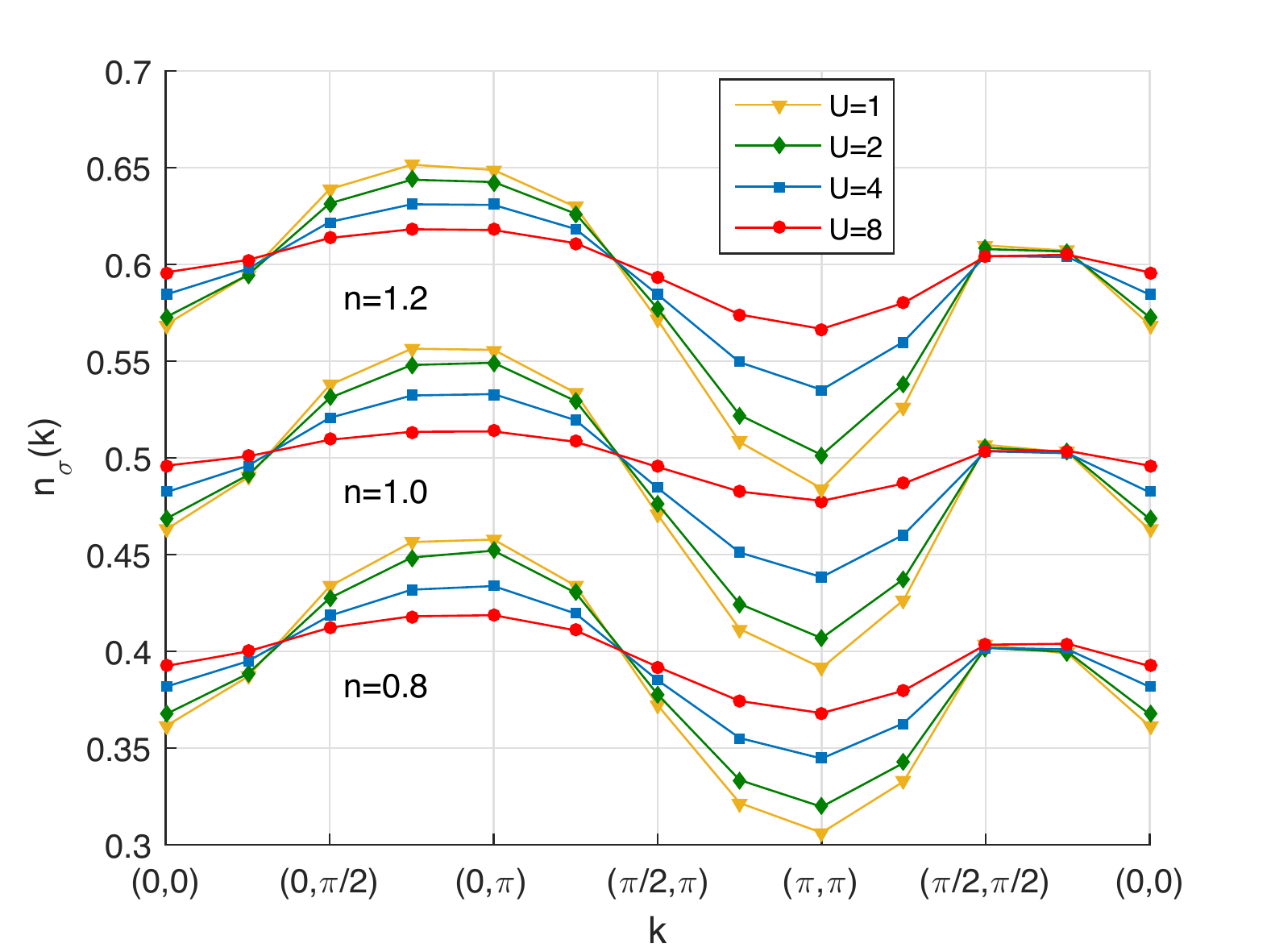}
\caption{\label{fig:MD-doped-approximation} (Color online) Momentum distribution for various doping and interaction parameters. Hard-band approximation. $n=0.8; 1.0; 1.2$, $\beta =1$} 
\end{figure}	

These results lead us to analyze the profiles of momentum distributions along the main crystallographic directions in hard-band approximation (Fig. \ref{fig:MD-doped-approximation}). It is possible in this situation to consider a greater level of doping at various values of $U$ away from the half-filling. As a result, Fig. \ref{fig:MD-doped-approximation} represents the momentum distribution of quasiparticle excitations of hole and electron types. Almost all the features of the momentum distribution demonstrated in Fig. \ref{fig:MD-doped} for the exact simulation are observed also in this case of the hard-band approximation. The distributions change little with the filling, and there is blurring of distribution gradients with the increase of the interaction. As for comparison of the profiles for various values of $U$, the following can be noted: in a first approximation, all the curves meet at points coinciding with the doping level, as in the half-filling case.

\section{Fermi-liquid parameters}
\label{7}
Studies of the Hubbard model \cite{24,25} and experimental data on FeAs-systems \cite{22} may indicate non-Fermi liquid nature of these strongly correlated systems. To investigate this issue, we extracted the Fermi-liquid parameters of the model \eqref{eq:Hamiltonian}. In the quasiparticle approximation, the spectral density near the maximum of a peak for electron and hole excitations can be described as follows \cite{26}:
\begin{equation}
A\left(\boldsymbol{k},\omega \right)=\frac{Z\left(\boldsymbol{k}\right)\Gamma\left(\boldsymbol{k},\omega \right)}{\pi \left({\left(\omega -\varepsilon \left(\boldsymbol{k}\right)+\mu \right)}^2+{\left(\Gamma \left(\boldsymbol{k},\omega \right)\right)}^2\right)}.
\end{equation}
Here $Z(\boldsymbol{k})$ is the quasiparticle weight or $Z$-factor, $\varepsilon \left(\boldsymbol{k}\right)$ is the excitation energy identified with the maximum of the spectral peak for a given point of the Brillouin zone (Fig. \ref{fig:DISP} (b), (d)), $\Gamma (\boldsymbol{k},\omega )$ is the quasiparticle scattering rate: 
\begin{equation}
\Gamma\left(\boldsymbol{k},\omega \right)=-Z\left(\boldsymbol{k}\right)\text{Im}\left(\Sigma\left(\boldsymbol{k},\omega \right)\right).
\end{equation}
Here $\Sigma\left(\boldsymbol{k},\omega \right)$ is the self-energy; it is assumed that the following condition is satisfied:
\begin{equation}
\left|(\varepsilon \left(\boldsymbol{k}\right)-\mu )\right|\gg \left|\Gamma \left(\omega=\mu;\boldsymbol{k}=k_F\right) \right|,
\end{equation}
so, the quasiparticle scattering rate is small.

$Z$-factor can also be obtained as follows [27]:
\begin{equation}\label{eq:Z-factor}
Z=\left(1-\frac{\partial (\text{Re}\Sigma \left(\omega\right)}{\partial \omega }\Big|_{\omega \to 0}\right)^{-1}.
\end{equation}
It is not possible to use the expression \eqref{eq:Z-factor} for $Z$-factor directly; one need to make an analytic continuation of the self-energy from imaginary to real frequencies and an extrapolation to zero temperature, using the fact that
\begin{equation}
\Sigma\left({\omega }_n\to 0^+\right)=\Sigma \left(\omega \to 0\right),
\end{equation}
where ${\omega }_n$ is the Matsubara frequency. This allows to rewrite \eqref{eq:Z-factor}:
\begin{equation}
Z=\frac{1}{1-\frac{\text{Im}\Sigma(\boldsymbol{k},i\omega _0)}{\omega _0}},
\end{equation}
where ${\omega }_0=\pi /\beta $ is zero fermion Matsubara frequency.

The relation between the Green's function and the self-energy is given by the Dyson equation (the chemical potential is set equal to zero):
\begin{equation}
G\left(\boldsymbol{k},i{\omega }_n\right)=\frac{1}{i{\omega }_n-\varepsilon \left(\boldsymbol{k}\right)-\Sigma(\boldsymbol{k},i{\omega }_n)}.
\end{equation}
Hence,
\begin{equation}\label{eq:Z-factor-final}
Z=\frac{\pi }{\beta }\frac{1}{\text{Im}\left[{\left(G\left(k,i{\omega }_0\right)\right)}^{-1}\right]}.
\end{equation}

\begin{figure}
\includegraphics[width=8cm]{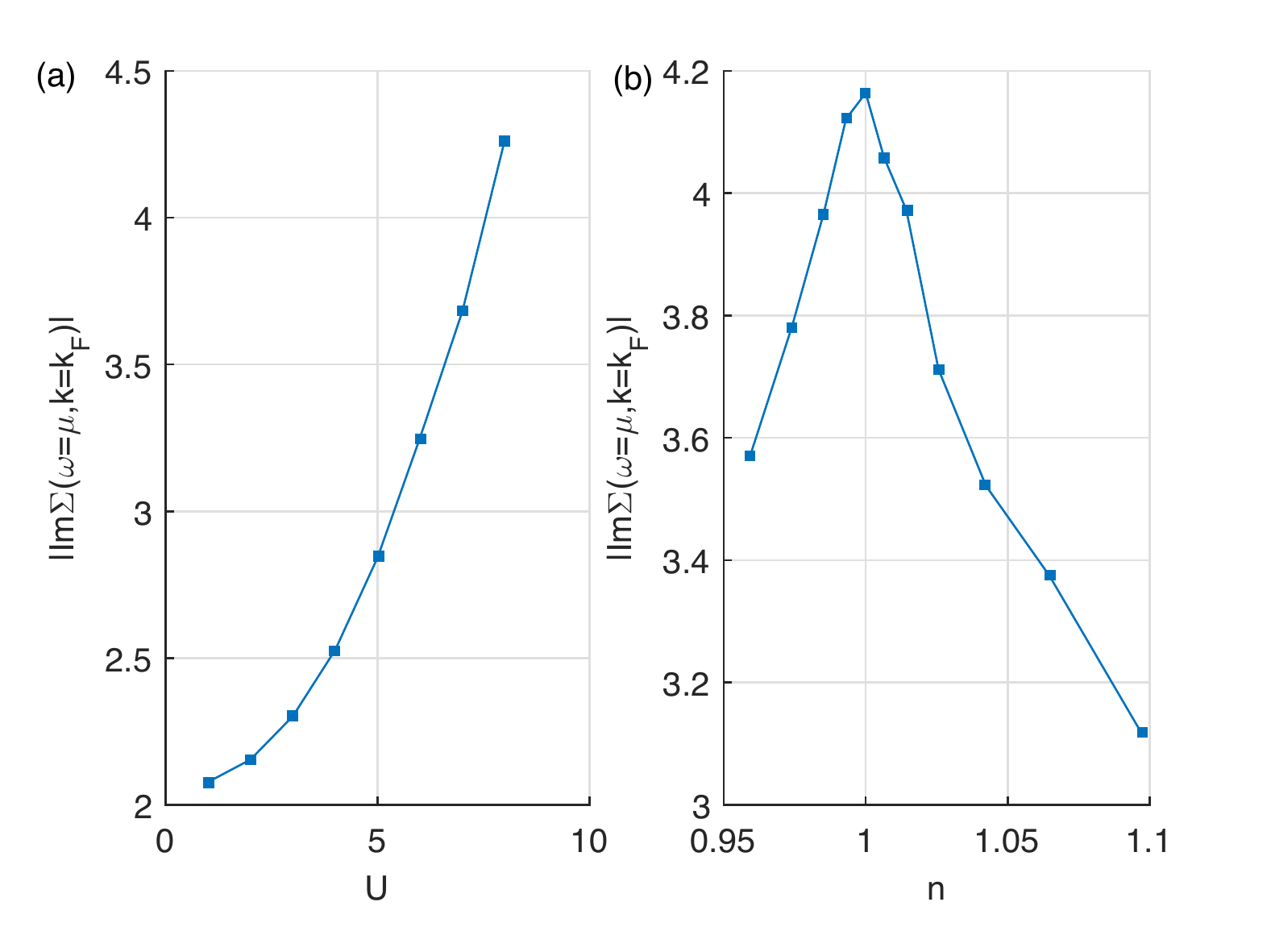}
\caption{\label{fig:IMSE} (Color online)  Imaginary part of the self-energy at $k=k_F$ as a function of a) interaction at $n=1$; b) doping at $U=4$} 
\end{figure}	

Figure \ref{fig:IMSE} shows the dependence of the self-energy on the strength of the interaction and the level of doping. We can notice that with increasing of $U$ the imaginary part of the self-energy grows as $U^2$; at the same time the damping of quasiparticles decreases: $\left|\left(\varepsilon \left(\boldsymbol{k}\right)-\mu \right)/\Gamma\left(\omega=\mu;\boldsymbol{k}=k_F\right)\right|\sim 1$ for $U\sim 1\div 2$ and $\left|\left(\varepsilon \left(\boldsymbol{k}\right)-\mu \right)/\Gamma\left(\omega=\mu;\boldsymbol{k}=k_F\right)\right|\sim 4$  for $U\sim 4\div 8$.

The dependence of imaginary part of the self-energy on doping is very close to that observed in \cite{27}. Electron-hole asymmetry is clearly visible; it can be explained by the asymmetry of Hamiltonian \eqref{eq:Hamiltonian}.

\begin{figure}
\includegraphics[width=8cm]{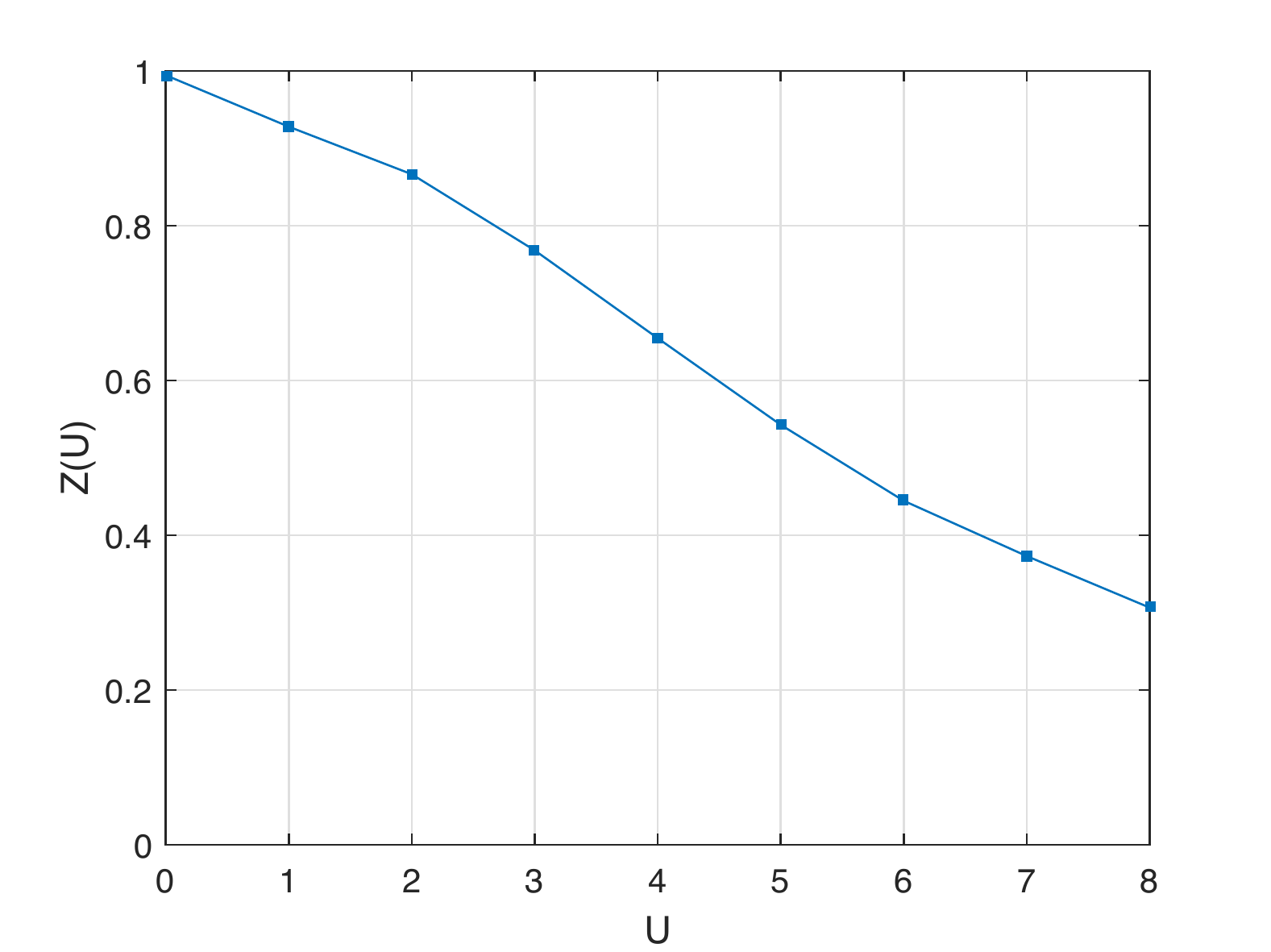}
\caption{\label{fig:Z-factor-U}  (Color online) $Z$-factor as function of $U$ at half-filling. $Z$-factor was calculated at $k=k_F$ according to \eqref{eq:Z-factor-final}} 
\end{figure}	

Figure \ref{fig:Z-factor-U} shows the dependence of $Z$-factor on the strength of the interaction. At $U=0$ clear Fermi-liquid behavior is observed. However, with the increase of the interaction the degree of non-Fermi-liquid behavior becomes more and more pronounced. It is important to notice that $Z$-factor and imaginary part of the self-energy are almost constant all over the first Brillouin zone (variation is less than 5\%).

\begin{figure}
\includegraphics[width=8cm]{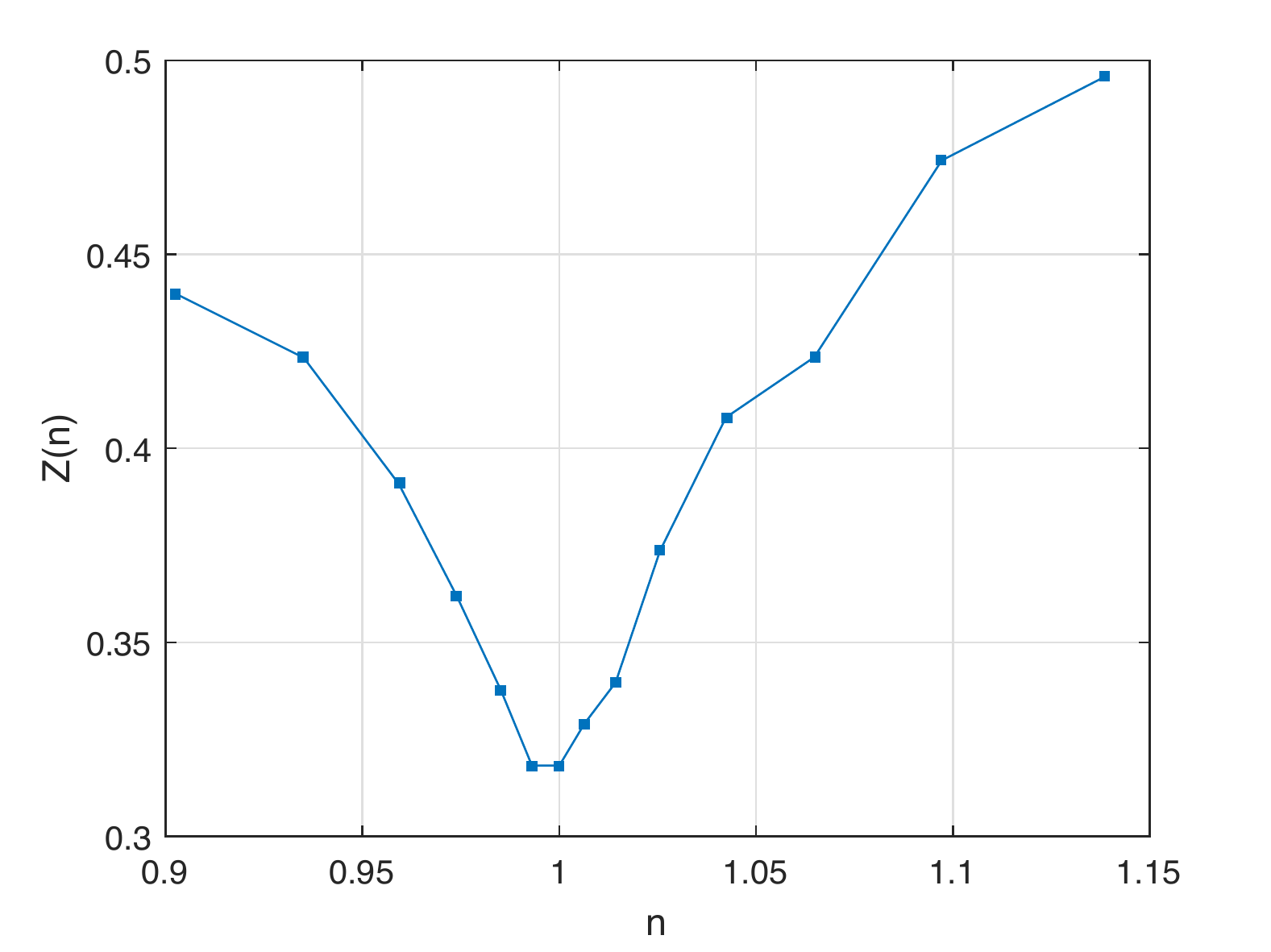}
\caption{\label{fig:Z-factor-n} (Color online) $Z$-factor as a function of doping. $U=4$; $\beta =2$} 
\end{figure}	

The effect of doping on $Z$-factor is shown at Fig. \ref{fig:Z-factor-n}. $Z$-factor increases significantly even at low doping; this indicates growth of DOS at the Fermi-level. Note that similar results were obtained in \cite{27} for single-band Hubbard model.

\begin{figure}
\includegraphics[width=8cm]{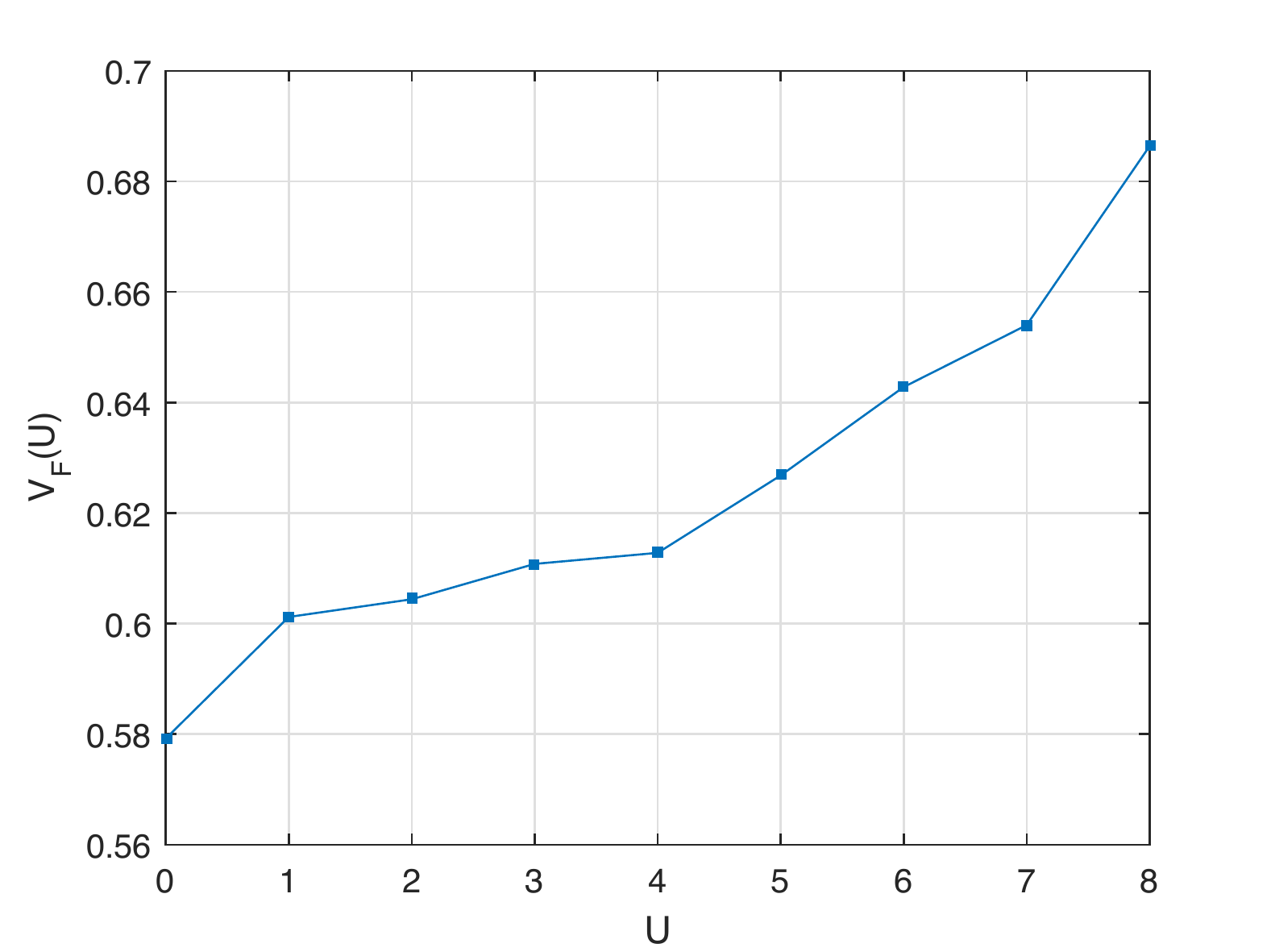}
\caption{\label{fig:FS-volume-U} (Color online) The volume of the FS as a function of $U$ at the half-filling} 
\end{figure}

At the end of this section, we compare the dependence of the volume of the FS on the interaction strength with predictions of Luttinger theorem (Fig. \ref{fig:FS-volume-U}). According to Luttinger theorem, the volume of the FS is independent on $U$ and changes linearly with doping. The evident deviation from the Luttinger theorem is caused presumably by two factors. First, as we previously noticed, $\beta =1$ is too large temperature to clearly define the FS. The temperature effect can be estimated from the difference $(V_F\left(U=0\right)-0.5)$. Second, the difference of the curve from the constant is due to the role of the interaction.

\section{Conclusions}
\label{Conclusions}
The two-dimensional two-orbital Hubbard model was studied with the use of finite-size cluster world line quantum Monte Carlo algorithm. 
This model is widely used for simulation of the band structure of FeAs clusters, which are structure elements of Fe-based high-temperature superconductors. 
We were able to overcome partly the sign problem and to obtain data for Matsubara Green's function at sufficiently low temperature in a wide range of model parameters. Spectral functions and the density of states were restored with the use of the stochastic optimization method in the undoped and low-doped regimes.
Profiles of the momentum distribution were obtained for the first Brillouin zone; they have pronounced jump near the Fermi level, which decreases with the growth of the strength of the interaction.
Fermi surface and its evolution were analysed using exact numerical calculations and within the framework of “hard-band” approximation, when simple filling of the bands is realized at constant DOS obtained from MC calculations in the undoped regime.
 It was shown, that the main features of the momentum distribution in the hard-band approximation are not much different from these of the exact calculation. 
Nesting in electronic and hole doped regimes was observed.
The Fermi-liquid parameters of the model were derived. 
$Z$-factor has shown continuous decreasing with growing of the strength of the interaction and sharp growth at the deviation from the half-filling. 
Electron-hole doping asymmetry of the $Z$-factor was revealed. 
The non-Fermi liquid behaviour and the deviation from Luttinger theorem were observed.

\section*{Acknowledgements}
The work was supported by Russian Science Foundation (project \# 16-19-00168).

\end{document}